\documentclass[aps,prl,twocolumn,final,floatfix,superscriptaddress]{revtex4-2}
\usepackage[utf8]{inputenc}
\usepackage{graphicx}
\usepackage{placeins}
\usepackage{floatflt}
\usepackage{epsfig,epsf,rotating}
\usepackage{amsmath}
\usepackage{amssymb}
\usepackage{natbib}
\usepackage{mathtools}
\usepackage{tikz}
\usepackage{ulem} 
\DeclareRobustCommand\sampleline[1]{%
  \tikz\draw[#1] (0,0) (0,\the\dimexpr\fontdimen22\textfont2\relax)
  -- (2em,\the\dimexpr\fontdimen22\textfont2\relax);%
}

\begin{document}
\bibliographystyle{apsrev4-2}

\title{Magnetic Confinement of an Ultracold Neutral Plasma}

\author{G.M. Gorman}
\affiliation{Rice University, Department of Physics and Astronomy, Houston, Texas, USA}
\author{M.K. Warrens}
\affiliation{Rice University, Department of Physics and Astronomy, Houston, Texas, USA}
\author{S.J. Bradshaw}
\affiliation{Rice University, Department of Physics and Astronomy, Houston, Texas, USA}
\author{T.C. Killian}
\affiliation{Rice University, Department of Physics and Astronomy, Houston, Texas, USA}

\date{\today}

\begin{abstract}
    We demonstrate magnetic confinement of an ultracold neutral plasma (UCNP) created at the null of a biconic cusp, or quadrupole magnetic field. Initially, the UCNP expands due to electron thermal pressure. As the plasma encounters stronger fields, expansion slows and the density distribution molds to the field. UCNP electrons  are strongly magnetized over most of the plasma, while ion magnetization is only significant at the boundaries. Observations suggest that electrons and ions  are predominantly trapped by magnetic mirroring and ambipolar electric fields respectively. Confinement times approach 0.5 ms, while unmagnetized plasmas dissipate on a timescale of a few tens of microseconds. 
\end{abstract}

\maketitle

\begin{figure}[h]
	\includegraphics[width=0.47\textwidth]{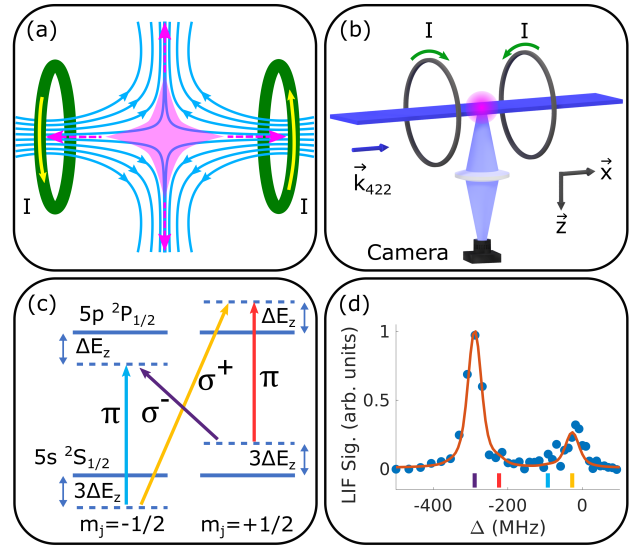}
	\caption{(a) Anti-Helmholtz coils, magnetic field lines and confined plasma in a biconic cusp field. Dashed arrows indicate loss gaps. (b) Experimental schematic for laser-induced fluorescence (LIF) imaging  and application of magnetic fields. The plasma is illuminated by a thin sheet of 422 nm light that propagates along the x-axis and is linearly polarized along the y-axis. The ion fluorescence is imaged onto an intensified CCD camera using a 1:1 optical relay along the z-axis.  (c) Sr levels involved in LIF with (\sampleline{dotted}) and without  (\sampleline{}) Zeeman shifts. (d) LIF spectrum of a magnetized UCNP at $x=4.6$ mm and $y=1.6$ mm (origin at plasma center), where  $B=70\,$G and the LIF laser drives the $\sigma$-transitions equally and the $\pi$-transitions are driven relatively weakly. The asymmetry in the spectrum reflects ion spin polarization.}
	\label{fig:exp-schem}
	\vspace{-.2in}
\end{figure}

The biconic cusp, or quadrupole magnetic field configuration  \cite{bgr58,spa71} formed by anti-Helmholtz current coils can confine neutral plasmas near the central null-field region
due to the magnetic-mirror effect \cite{pef60}. This confinement scheme has been of long-standing interest, initially for magnetic-confinement fusion \cite{bgr58,spa71,hai77,kts74,lhm76}, and more recently for ion sources for applications such as material processing and ion thrusters \cite{cgc11,cwk16,hbw14}. Neutral plasma expanding across biconic cusp field lines experiences changing length scales and dominant physical processes, and the complex geometry has similarity to the solar wind interacting with the Earth's magnetosphere \cite{spa71,rus00}. Here, we demonstrate the magnetic confinement of an ultracold neutral plasma (UCNP) \cite{kkb99,kpp07,lro17} created at the null point of a biconic cusp field (Fig. \ref{fig:exp-schem}(a)). 

UCNPs, created here by photoionizing laser-cooled Sr atoms near the ionization threshold, have ion temperatures $T_i\sim 1$\,K and tunable electron temperatures of $T_e=1-1000$\,K, which offers a novel regime for study of magnetized and magnetically confined neutral plasmas. UCNPs also provide the opportunity to study the combined effects of
 magnetization and strong coupling on collisional and transport phenomena because ions are strongly coupled in UCNPs, with the ratio of Coulomb energy to kinetic energy, known as the Coulomb coupling parameter \cite{ich82}, as high as $\Gamma_i=11$ \cite{scg04,lbm13,lgk19}. Electrons can also approach the strongly coupled regime, with $\Gamma_e \lesssim 0.4$ \cite{kon02,mck02,rha02,gls07,cwr17}. There is emerging focus on magnetized and strongly coupled plasmas in general  \cite{obo11,bda17a} and in the ultracold regime  \cite{bda17a,iga18,tba18}, driven in large part by new experimental capabilities in dusty  \cite{tmr12,kdd19,flw19} and laser-produced high-energy-density plasmas \cite{sqf18,sbe18}.

Previous experimental work with magnetized UCNPs is limited. A pioneering experiment studied cross-field ambipolar diffusion in a uniform field strong enough to magnetize electrons but not ions \cite{zfr08expansion}. The same authors \cite{zfr08} identified a high-frequency electron drift instability in weak, crossed magnetic and electric fields. Non-neutral plasmas are routinely confined in combined electric and magnetic fields in Penning-Malmberg traps \cite{don99}, and these techniques have been extended to confine partially overlapping clouds of positive and negative charges at ultracold temperatures in nested traps, such as for  anti-hydrogen production \cite{aab02,gbo02}. Dynamics of a UCNP loaded into such a trap, forming partially overlapping electron and ion components,  was studied in \cite{ckz08}.

To create a UCNP, Sr atoms are first laser-cooled and confined with a magneto-optical trap (MOT) using the $5s^2\prescript{1}{}\!S_0 - 5s5p\prescript{1}{}\!P_1$ transition at 461 nm. Atoms populate  the metastable $5s5p \prescript{3}{}\!P_{2}$ state throughout the laser cooling process due to a weak decay path from the cycling transition, and low-field seeking  $\prescript{3}{}\!P_{2}$ atoms ($m_j=+2,+1$) are trapped in the quadrupole  magnetic field of the MOT due to their large magnetic moment \cite{nsl03}.  $\prescript{3}{}\!P_{2}$ atoms are then photoionized  near threshold with  322\,nm photons from a 10-ns pulsed dye laser.  The plasma inherits its initial density distribution, $n(\vec{r}) = n_0\,\mathrm{exp}( -{\sqrt{x^2+ {y^2}/{4} + {z^2}/{4}}}/{\alpha})$,  from  the magnetically  trapped neutral atoms, with $\alpha={3k_BT_a}/{8\mu_B B'} \approx 1$\,mm, where $T_a\approx 3$ mK is the temperature of the neutral atoms, $B'=150$\,G/cm is the gradient of the magnetic field along the symmetry axis, and $\mu_B$ and $k_B$ are the Bohr magneton and Boltzmann constant, respectively.  The atom cloud is small compared to the radius of the coils creating the magnetic field, so a linear approximation of the field profile is sufficient. 

Initial peak density ($n_0\sim 10^{9}$\,cm$^{-3}$) is controlled by varying the  laser-cooling time  ($\sim 1$\,s) to vary the number of trapped atoms.
Dynamics with or without magnetic field are studied by leaving the field on or extinguishing it prior to photoionization. The initial  electron  temperature, $T_{e}=20-160$ K, is set by the photoionization laser detuning above threshhold. Ions are created with extremely low kinetic energy, close to that of the precursor neutral atoms, but they possess significant potential energy due to their initially uncorrelated state and  undergo a process called disorder-induced heating in the first few 100 ns. This results in ion temperatures $T_i\sim1\,$K  \cite{mur01,scg04}.

The plasma is probed at an adjustable time after photoionization using laser-induced fluorescence (LIF) on the $5s\prescript{2}{}\!S_{1/2}-5p\prescript{2}{}\!P_{1/2}$ transition of Sr$^+$ at 422 nm \cite{cgk08}. The LIF laser, with detuning $\Delta$ from unperturbed resonance, illuminates a 1-mm-thick central slice of the plasma ($z\approx0$)  (Fig. \ref{fig:exp-schem}(b)). Scattered photons are imaged onto an intensified CCD camera to obtain a spatially resolved LIF fluorescence spectrum, $F(x,y,\Delta)$, with $50\,\mu$m  resolution. Each image consists of photons collected over a $0.5-2\,\mu$s time period that is set by CCD and LIF-laser gating. Image times indicated in all figures refer to the midway point of the imaging window.

The local ion density,  temperature, and hydrodynamic fluid velocity along the LIF laser propagation direction ($v_{x,hyd}$)  are extracted from $F(x,y,\Delta)$ with a Fermi's golden rule model of the LIF spectrum.  In the absence of magnetic fields, the levels involved in LIF are degenerate. For magnetized UCNPs, however, the levels are Zeeman-shifted (Fig. \ref{fig:exp-schem}(c)), and the strengths of different Zeeman components reflect the initial spin polarization of the ions and decomposition of the LIF laser polarization in the coordinate frame of the local magnetic field. Doppler broadenings due to thermal motion and shifts due to hydrodynamic fluid velocity are incorporated by convolving the line strengths with the local ion velocity distribution, assuming local thermal equilibrium. Fig. \ref{fig:exp-schem}(d) shows an example of such a fit.  Asymmetric spectra are observed consistently across the plasma spatially and during the entire plasma evolution, indicating long-lived polarization of the spin of the valence electron of each Sr$^+$ ion along the local magnetic field. Polarization arises because the precursor magnetically trapped atoms posses a high degree of spin polarization.

\begin{figure*}[ht!]
    \centering
    \includegraphics[width=.95\textwidth]{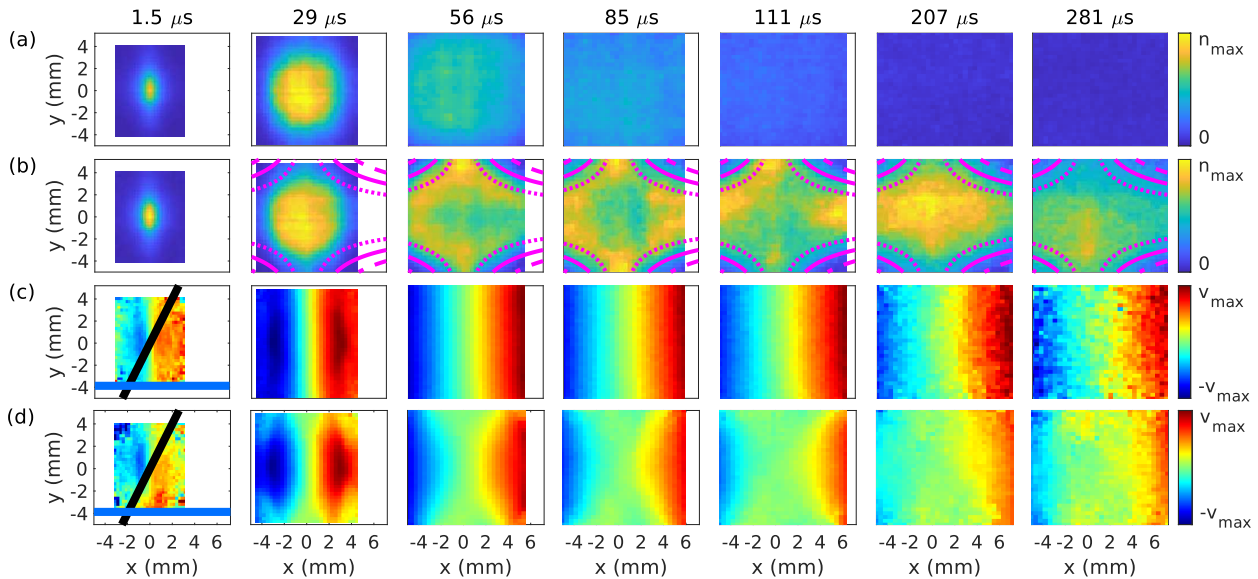}
    \caption{
   Plasma expansion without and with  magnetic field ($B'=$ 150 G/cm)  for $T_e (0)=20$ K.  Rows (a) and  (b) show density distributions without and with field, respectively. The lines in row B (\sampleline{dotted},  \sampleline{},  \sampleline{dashed}), correspond to sample field lines. Time after plasma creation is indicated above each column. The scale for the density color bar for each time point is $n_{max}=[13.8, 1.7, 0.59, 0.36, 0.25,0.18,0.18]\times10^{8}$  cm$^{-3}$, in order of increasing  time. Rows (c) and  (d) show the x-component of the hydrodynamic velocity, $v_{x,hyd}$,  without and with  magnetic fields, respectively. The scale for the velocity color bar 
    for each time point is $v_{max}=[10, 70, 100, 75, 62, 35, 25]$ m/s. The black and blue lines in the first time point of rows c and d correspond to $y=2x$ and $y = -5$ mm, respectively. Fig. 3 plots  $v_{x,hyd}$  along each of these lines.   
  }
  \label{fig:n-v-distro}
\end{figure*}

In a UCNP, when magnetic forces are absent or negligible, electron thermal pressure drives plasma expansion on a characteristic hydrodynamic timescale of $\tau_{exp}=\sqrt{{m_i\sigma(0)^2}/{k_BT_e(0)}}$, where $m_i$ is the Sr$^+$ mass  and $\sigma(0)$ and $T_e(0)$ are the initial geometric mean RMS plasma radius and electron temperature, respectively \cite{kkb00,rha02}.  This phenomenon has traditionally been studied with UCNPs with spherically symmetric Gaussian density distributions \cite{lgs07}, but the general description  applies reasonably well for exponentially decaying density distributions as used here \cite{wgk20}. For Guassian plasmas, the expansion velocity field increases with  distance from plasma center and time after photoionization ($t$), following $\vec{v}_{hyd}=\gamma(t)\vec{r}$ where $\gamma(t)=t/[\tau_{exp}^2(1+t^2/\tau_{exp}^2)]$ \cite{bku98,dse98,lgs07}.

Fig. \ref{fig:n-v-distro} shows the evolution of plasma density and  $v_{x,hyd}$ without and with magnetic fields for a UCNP with $T_e(0)=20$ K and $\sigma(0)=1.3$ mm ($\tau_{exp}=30\,\mu$s). Initially ($t=1.5 \,\mu$s), the plasma density and velocity distributions without (Fig. \ref{fig:n-v-distro}(a) and (c)) and with (Fig. \ref{fig:n-v-distro}(b) and (d)) the fields are nearly identical and the plasma has not yet developed significant expansion velocity. 

\begin{figure}[b!]
	\includegraphics[width=0.47\textwidth]{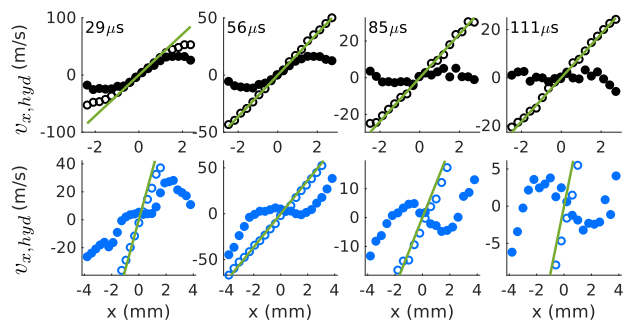}
	\caption{
		Evolution of the x-component of hydrodynamic velocity for plasma along the lines $y=2x$ (top) and $y=-5$ mm (bottom) for data shown in Fig. \ref{fig:n-v-distro}  for plasmas evolving without (open symbols) and with (closed symbols) magnetic fields. The solid green line represents $v_{x,hyd}=x/t$, the ion velocity  expected for $\vec{v}_{hyd}=\gamma(t)\vec{r}$ or ballistic expansion at late times ($t\gg \tau_{exp}$). Time since plasma creation is indicated above each column. 
	}
	\label{fig:v-trans}
\end{figure}

At early times ($t<\tau_{exp}$), outward hydrodynamic pressure dominates and the expansion is relatively unaffected by the magnetic fields. This is reflected by the nearly identical density distributions at $t=29\,\mu$s. However, while constant $v_{x,hyd}$ along lines of constant $x$ is observed without fields as expected for $\vec{v}_{hyd}\propto \vec{r}$,  velocity retardation is evident for ions traversing field lines  in regions far from plasma center where the fields are large. For example, along the $y=2x$ and $y=-5$ mm lines (indicated in Fig.\ \ref{fig:n-v-distro}(c) and (d), $t=1.5\,\mu$s),  magnetic forces have clearly impeded the expansion by $t=29\,\mu$s. This is highlighted in the velocity transects shown in Fig.\ \ref{fig:v-trans}. Along the $x$-axis, where plasma expansion velocity is parallel to the field lines, $v_{x,hyd}$ is still unaffected by fields.

By $t=85\,\mu$s the no-fields plasma appears  uniform because the plasma size exceeds the imaged region. Gradients in plasma density that produce outward electron thermal pressure have diminished, leading to a ballistic  plasma expansion for no fields that persists throughout the rest of the expansion (green line in Fig. \ref{fig:v-trans} representing $v_{x,hyd}=x/t$). 

The with-fields plasma shows considerable effects of the magnetic fields during this intermediate time. The plasma begins to develop a boundary that follows a field line (Fig.\ \ref{fig:n-v-distro}(b)).   Expansion near the plasma center is relatively unaffected because of the weak fields, leading to a central plasma depletion and a build up of plasma density at the boundary. The effects on the plasma velocity, evident at earlier times, become stronger. Most striking is a reversal of $v_{x,hyd}$ in regions of strong field where the expansion velocity is close to perpendicular to field lines, such as near $x\pm 2\,$mm for $y=-5$\,mm (Fig.\ \ref{fig:v-trans}(bottom)). The combined density and velocity information indicates that plasma flow  across field lines in these regions has halted, and is now redirected along field lines back towards $x=0$ and increasing $|y|$ for $|x|\lesssim 2$\,mm.

The initial  expansion of the UCNP is driven by gradients in the hydrodynamic electron thermal pressure \cite{kpp07}, which creates a force per ion of
\begin{equation} \label{Eq:ExpansionForce}
    \vec{F}_{exp}(\vec{r},t)=-\frac{\nabla n(\vec{r})}{n(\vec{r})}k_{B}T_{e}(t),
\end{equation}
where $T_{e}(t)$ is the  electron temperature, which decreases during expansion due to adiabatic cooling. Deviations between no-fields and with-fields plasma velocity for ions crossing field lines are observed as early as $t=29\,\mu$s (for example, along the velocity transects in Fig. \ref{fig:v-trans}). The Lorentz force that the fields exert on charged particles of species $s$ due to the expansion velocity, $\vec{F}_{L}(\vec{r},t)=q_s\,\vec{v}_{x,hyd}(\vec{r},t)\times \vec{B}(\vec{r})$ is negligible compared to $\vec{F}_{exp}$ at this time, and is not responsible for the velocity reduction.

However, at the boundary of the with-fields plasma, where the effects of the magnetic field on the expansion first appear, the magnetization of the plasma is significant, as characterized by the  magnetization parameter $\delta=\rho/L$,  the ratio of thermal gyroradius $\rho=\sqrt{k_B T_s m_s}/eB$ to the characteristic lengthscale of the plasma ${n(\vec{r})}/{\nabla n(\vec{r})}\equiv L\sim 1$\,mm. For $B\approx50$\,G and $T_e=10$\,K, $\rho_e =10\,\mu$m, indicating the electrons are strongly magnetized. For ions with $T_i=0.25$\,K,   the magnetization is modest, with $\rho_i \approx L$. This suggests that the reduction of cross-field expansion results from pinning of the strongly magnetized electrons to field lines. Assuming classical diffusion \cite{zfr08expansion},  the transverse diffusion constant is $D_{\perp}=\rho_e^2 \nu_{ei}\approx 1 \,\mathrm{mm}^2/\mathrm{ms}$, for ion density $n_i=2\times 10^{7}$\,cm$^{-3}$, where $\nu_{ei}\propto n_i/T_e^{3/2}$ is the electron-ion collision rate. Cross-field diffusion several mm from the field null should thus be slow on the timescale of these experiments, as observed. 

At late  times (Fig.\ \ref{fig:n-v-distro}, $t> 150\,\mu\mathrm{s}\sim 5\tau_{exp}$), the expansion for the magnetized plasma has essentially halted and the plasma has become magnetically trapped, with a density maximum in the plasma center. Magnetic confinement arises from the magnetic mirror effect, which results from conservation of the adiabatic invariant $\mu\equiv  m_{s} v_{\perp}^2/2B$ for a charged particle moving along a guiding field line,  where $v_{\perp}$ is the particle's velocity transverse to the local field. The force per particle is $\vec{F}_{\|}=-\hat{\ell}\mu \partial B/\partial \ell$ pointing in the direction of decreasing field strength, where $\ell$ is the distance along the field line. Particles with large enough $\mu$   and small enough total kinetic energy $\mathcal{E}$, described as having a large pitch angle, are trapped between bounce points where $B=\mathcal{E}/\mu$.

Ion motion for plasma regions imaged in Fig.\ \ref{fig:n-v-distro} is not adiabatic, but electron motion is adiabatic everywhere except close to the field null. Thus, as expected for  biconic cusp fields \cite{bgr58,spa71,lhm76,bme86c}, the likely description of UCNP trapping is that electrons move along field lines, confined between bounce points. Ambipolar electric fields transverse to magnetic field lines constrain ion cross-field  transport.
Electrons with large enough $\mathcal{E}/\mu$  escape through loss gaps around the field maxima along field lines in the $x-y$ plane and along the $z$ symmetry axis (Fig.\ \ref{fig:exp-schem}(a)).  This loss is fed by non-adiabatic mixing of electron trajectories near the field null and by collisions. In the loss gaps, ambipolar fields slow electrons, leading to plasma loss at ion acoustic velocities $\propto \sqrt{T_e/m_i}$ \cite{bme86c}. 

\begin{figure}
	\includegraphics[width=0.47\textwidth]{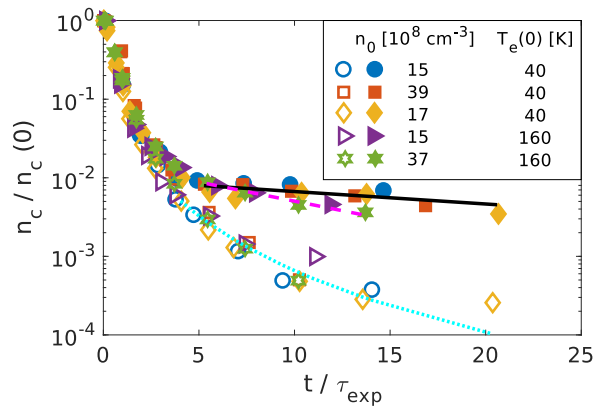}
	\caption{
			Relative plasma density in the plasma center versus time scaled by the hydrodynamic expansion timescale ($\tau_{exp}$)  without (open symbols) and with (closed symbols) magnetic field. Solid and dashed lines are exponential fits to $T_e(0)=40$ K and $T_e(0)=160$ K with-fields data, respectively, for $t>5\tau_{exp}$. The dotted line is a $1/t^3$  fit to all no-fields data with $t>4\tau_{exp}$. Initial peak density and electron temperature are indicated in the legend.
	}
	\label{fig:n-center}
\end{figure}

To characterize the onset of confinement and rate of plasma loss, Fig. \ref{fig:n-center} shows the time evolution of the central plasma density $n_c(t)$ relative to its initial value. Time is scaled by the characteristic hydrodynamic timescale. A $1/t^3$ fit to data with $t>4\tau_{exp}$ demonstrates the ballistic nature of plasma expansion without fields at late times.

In contrast, the with-fields (closed symbol) central density  stabilizes at $n_c(t)/n_c(0)= 10^{-2}$ at  $t= 5\tau_{exp}$. Developing a quantitative explanation for this universal behavior will the subject of future study, but it is generally consistent with the onset of trapping occurring when the average hydrodynamic expansion force (Eq.\ \ref{Eq:ExpansionForce}), which decreases rapidly with time, matches the typical magnetic mirror force. While the onset time and level of trapping in scaled units shows no discernible dependence on density, the trap lifetime decreases for higher $T_e(0)$, which is consistent with plasma loss through loss gaps at the ion acoustic speed. The solid and dashed lines in Fig. \ref{fig:n-center} are exponential fits to $T_e=40\,$K and $T_e=160\,$K data, respectively, which reveal magnetic confinement timescales of $500\,\mu$s (40 K) and $100\,\mu$s (160 K). 

This work demonstrates magnetic confinement of a UCNP in a biconic cusp magnetic field. The plasma density and velocity-field profiles and estimates of relevant forces
imply that the plasma confinement results from strongly magnetized electrons following guiding field lines and ions constrained by ambipolar fields.
Observed trap lifetimes decrease with increasing electron temperature, which is consistent with a dominant loss mechanism of flux through the loss gaps.

The magnetic confinement of UCNPs opens many new  research directions. UCNPs have long been used for experimental studies of the effects of strong coupling on collisional transport processes \cite{bbl19}, and these new capabilities may enable exploration of overlapping regimes of strong coupling and magnetization \cite{obo11,bda17a}.  With experimental improvements such as increased magnetic field gradient and field of view for LIF imaging, it should be possible to characterize scaling of trapping behavior with magnetic field and study plasma flow in loss gaps, which will support development of a quantitative model of plasma dynamics. The combination of magnetic trapping with recently demonstrated techniques of laser cooling of UCNP ions \cite{lgk19} appears promising for improving laser cooling efficacy. Laser-induced forces also offer a new tool for plugging loss gaps in a biconic cusp trap, perhaps leading to significantly enhanced trap lifetimes.

This work was supported by the Air Force Office of Scientific Research through Grant No. FA9550-17-1-0391 and the National Science Foundation Graduate Research Fellowship Program under Grant No. 1842494. 


\begin{thebibliography}{47}%
	\makeatletter
	\providecommand \@ifxundefined [1]{%
		\@ifx{#1\undefined}
	}%
	\providecommand \@ifnum [1]{%
		\ifnum #1\expandafter \@firstoftwo
		\else \expandafter \@secondoftwo
		\fi
	}%
	\providecommand \@ifx [1]{%
		\ifx #1\expandafter \@firstoftwo
		\else \expandafter \@secondoftwo
		\fi
	}%
	\providecommand \natexlab [1]{#1}%
	\providecommand \enquote  [1]{``#1''}%
	\providecommand \bibnamefont  [1]{#1}%
	\providecommand \bibfnamefont [1]{#1}%
	\providecommand \citenamefont [1]{#1}%
	\providecommand \href@noop [0]{\@secondoftwo}%
	\providecommand \href [0]{\begingroup \@sanitize@url \@href}%
	\providecommand \@href[1]{\@@startlink{#1}\@@href}%
	\providecommand \@@href[1]{\endgroup#1\@@endlink}%
	\providecommand \@sanitize@url [0]{\catcode `\\12\catcode `\$12\catcode
		`\&12\catcode `\#12\catcode `\^12\catcode `\_12\catcode `\%12\relax}%
	\providecommand \@@startlink[1]{}%
	\providecommand \@@endlink[0]{}%
	\providecommand \url  [0]{\begingroup\@sanitize@url \@url }%
	\providecommand \@url [1]{\endgroup\@href {#1}{\urlprefix }}%
	\providecommand \urlprefix  [0]{URL }%
	\providecommand \Eprint [0]{\href }%
	\providecommand \doibase [0]{https://doi.org/}%
	\providecommand \selectlanguage [0]{\@gobble}%
	\providecommand \bibinfo  [0]{\@secondoftwo}%
	\providecommand \bibfield  [0]{\@secondoftwo}%
	\providecommand \translation [1]{[#1]}%
	\providecommand \BibitemOpen [0]{}%
	\providecommand \bibitemStop [0]{}%
	\providecommand \bibitemNoStop [0]{.\EOS\space}%
	\providecommand \EOS [0]{\spacefactor3000\relax}%
	\providecommand \BibitemShut  [1]{\csname bibitem#1\endcsname}%
	\let\auto@bib@innerbib\@empty
	\bibitem [{\citenamefont {Berkowitz}\ \emph {et~al.}(1958)\citenamefont
		{Berkowitz}, \citenamefont {Grad},\ and\ \citenamefont {Rubin}}]{bgr58}%
	\BibitemOpen
	\bibfield  {author} {\bibinfo {author} {\bibfnamefont {J.}~\bibnamefont
			{Berkowitz}}, \bibinfo {author} {\bibfnamefont {H.}~\bibnamefont {Grad}},\
		and\ \bibinfo {author} {\bibfnamefont {H.}~\bibnamefont {Rubin}},\ }in\
	\href@noop {} {{\bibinfo {booktitle} {\textit{Proceedings of the 2nd
				International Conference on Peaceful Uses of Atomic Energy}}}},\ Vol.~\bibinfo
	{volume} {1}\ (\bibinfo {address} {Geneva, Switzerland},\ \bibinfo {year}
	{1958})\ p.\ \bibinfo {pages} {177}\BibitemShut {NoStop}%
	\bibitem [{\citenamefont {Spalding}(1971)}]{spa71}%
	\BibitemOpen
	\bibfield  {author} {\bibinfo {author} {\bibfnamefont {I.}~\bibnamefont
			{Spalding}},\ }in\ \href@noop {} {{\bibinfo {booktitle} {\textit{Advances in
				Plasma Physics}}}},\ Vol.~\bibinfo {volume} {4}\ (\bibinfo  {publisher} {Ed.
		by A. Simon and W. B. Thompson. Interscience},\ \bibinfo {address} {New
		York},\ \bibinfo {year} {1971})\ p.~\bibinfo {pages} {79}\BibitemShut
	{NoStop}%
	\bibitem [{\citenamefont {Post}\ \emph {et~al.}(1960)\citenamefont {Post},
		\citenamefont {Ellis}, \citenamefont {Ford},\ and\ \citenamefont
		{Rosenbluth}}]{pef60}%
	\BibitemOpen
	\bibfield  {author} {\bibinfo {author} {\bibfnamefont {R.~F.}\ \bibnamefont
			{Post}}, \bibinfo {author} {\bibfnamefont {R.~E.}\ \bibnamefont {Ellis}},
		\bibinfo {author} {\bibfnamefont {F.~C.}\ \bibnamefont {Ford}},\ and\
		\bibinfo {author} {\bibfnamefont {M.~N.}\ \bibnamefont {Rosenbluth}},\ }\href
	{https://doi.org/10.1103/PhysRevLett.4.166} {\bibfield  {journal} {\bibinfo
			{journal} {Phys. Rev. Lett.}\ }\textbf {\bibinfo {volume} {4}},\ \bibinfo
		{pages} {166} (\bibinfo {year} {1960})}\BibitemShut {NoStop}%
	\bibitem [{\citenamefont {Haines}(1977)}]{hai77}%
	\BibitemOpen
	\bibfield  {author} {\bibinfo {author} {\bibfnamefont {M.~G.}\ \bibnamefont
			{Haines}},\ }\href {https://doi.org/10.1088/0029-5515/17/4/015} {\bibfield
		{journal} {\bibinfo  {journal} {Nucl. Fusion}\ }\textbf {\bibinfo {volume}
			{17}},\ \bibinfo {pages} {811} (\bibinfo {year} {1977})}\BibitemShut
	{NoStop}%
	\bibitem [{\citenamefont {Kitsunezaki}\ \emph {et~al.}(1974)\citenamefont
		{Kitsunezaki}, \citenamefont {Tanimoto},\ and\ \citenamefont
		{Sekiguchi}}]{kts74}%
	\BibitemOpen
	\bibfield  {author} {\bibinfo {author} {\bibfnamefont {A.}~\bibnamefont
			{Kitsunezaki}}, \bibinfo {author} {\bibfnamefont {M.}~\bibnamefont
			{Tanimoto}},\ and\ \bibinfo {author} {\bibfnamefont {T.}~\bibnamefont
			{Sekiguchi}},\ }\href {https://doi.org/10.1063/1.1694636} {\bibfield
		{journal} {\bibinfo  {journal} {Phys. Fluids}\ }\textbf {\bibinfo {volume}
			{17}},\ \bibinfo {pages} {1895.} (\bibinfo {year} {1974})}\BibitemShut
	{NoStop}%
	\bibitem [{\citenamefont {Leung}\ \emph {et~al.}(1976)\citenamefont {Leung},
		\citenamefont {Hershkowitz},\ and\ \citenamefont {MacKenzie}}]{lhm76}%
	\BibitemOpen
	\bibfield  {author} {\bibinfo {author} {\bibfnamefont {K.~N.}\ \bibnamefont
			{Leung}}, \bibinfo {author} {\bibfnamefont {N.}~\bibnamefont {Hershkowitz}},\
		and\ \bibinfo {author} {\bibfnamefont {K.~R.}\ \bibnamefont {MacKenzie}},\
	}\href {https://doi.org/10.1063/1.861575} {\bibfield  {journal} {\bibinfo
			{journal} {Phys. Fluids}\ }\textbf {\bibinfo {volume} {19}},\ \bibinfo
		{pages} {1045} (\bibinfo {year} {1976})}\BibitemShut {NoStop}%
	\bibitem [{\citenamefont {Carr}\ \emph {et~al.}(2011)\citenamefont {Carr},
		\citenamefont {Gummersall}, \citenamefont {Cornish},\ and\ \citenamefont
		{Khachan}}]{cgc11}%
	\BibitemOpen
	\bibfield  {author} {\bibinfo {author} {\bibfnamefont {M.}~\bibnamefont
			{Carr}}, \bibinfo {author} {\bibfnamefont {D.}~\bibnamefont {Gummersall}},
		\bibinfo {author} {\bibfnamefont {S.}~\bibnamefont {Cornish}},\ and\ \bibinfo
		{author} {\bibfnamefont {J.}~\bibnamefont {Khachan}},\ }\href
	{https://doi.org/10.1063/1.3655446} {\bibfield  {journal} {\bibinfo
			{journal} {Phys. Plasmas}\ }\textbf {\bibinfo {volume} {18}},\ \bibinfo
		{pages} {112501} (\bibinfo {year} {2011})}\BibitemShut {NoStop}%
	\bibitem [{\citenamefont {Cooper}\ \emph {et~al.}(2016)\citenamefont {Cooper},
		\citenamefont {Weisberg}, \citenamefont {Khalzov}, \citenamefont {Milhone},
		\citenamefont {Flanagan}, \citenamefont {Peterson}, \citenamefont {Wahl},\
		and\ \citenamefont {Forest}}]{cwk16}%
	\BibitemOpen
	\bibfield  {author} {\bibinfo {author} {\bibfnamefont {C.~M.}\ \bibnamefont
			{Cooper}}, \bibinfo {author} {\bibfnamefont {D.~B.}\ \bibnamefont
			{Weisberg}}, \bibinfo {author} {\bibfnamefont {I.}~\bibnamefont {Khalzov}},
		\bibinfo {author} {\bibfnamefont {J.}~\bibnamefont {Milhone}}, \bibinfo
		{author} {\bibfnamefont {K.}~\bibnamefont {Flanagan}}, \bibinfo {author}
		{\bibfnamefont {E.}~\bibnamefont {Peterson}}, \bibinfo {author}
		{\bibfnamefont {C.}~\bibnamefont {Wahl}},\ and\ \bibinfo {author}
		{\bibfnamefont {C.~B.}\ \bibnamefont {Forest}},\ }\href
	{https://doi.org/10.1063/1.4963850} {\bibfield  {journal} {\bibinfo
			{journal} {Phys. Plasmas}\ }\textbf {\bibinfo {volume} {23}},\ \bibinfo
		{pages} {102505} (\bibinfo {year} {2016})}\BibitemShut {NoStop}%
	\bibitem [{\citenamefont {Hubble}\ \emph {et~al.}(2014)\citenamefont {Hubble},
		\citenamefont {Barnat}, \citenamefont {Weatherford},\ and\ \citenamefont
		{Foster}}]{hbw14}%
	\BibitemOpen
	\bibfield  {author} {\bibinfo {author} {\bibfnamefont {A.~A.}\ \bibnamefont
			{Hubble}}, \bibinfo {author} {\bibfnamefont {E.~V.}\ \bibnamefont {Barnat}},
		\bibinfo {author} {\bibfnamefont {B.~R.}\ \bibnamefont {Weatherford}},\ and\
		\bibinfo {author} {\bibfnamefont {J.~E.}\ \bibnamefont {Foster}},\ }\href
	{https://doi.org/10.1088/0963-0252/23/2/022001} {\bibfield  {journal}
		{\bibinfo  {journal} {Plasma Sources Sci. and Technol.}\ }\textbf {\bibinfo
			{volume} {23}},\ \bibinfo {pages} {022001} (\bibinfo {year}
		{2014})}\BibitemShut {NoStop}%
	\bibitem [{\citenamefont {Russell}(2000)}]{rus00}%
	\BibitemOpen
	\bibfield  {author} {\bibinfo {author} {\bibfnamefont {C.~T.}\ \bibnamefont
			{Russell}},\ }\href {https://doi.org/10.1109/27.902211} {\bibfield  {journal}
		{\bibinfo  {journal} {IEEE Trans. Plasma Sci.}\ }\textbf {\bibinfo {volume}
			{28}},\ \bibinfo {pages} {1818} (\bibinfo {year} {2000})}\BibitemShut
	{NoStop}%
	\bibitem [{\citenamefont {Killian}\ \emph {et~al.}(1999)\citenamefont
		{Killian}, \citenamefont {Kulin}, \citenamefont {Bergeson}, \citenamefont
		{Orozco}, \citenamefont {Orzel},\ and\ \citenamefont {Rolston}}]{kkb99}%
	\BibitemOpen
	\bibfield  {author} {\bibinfo {author} {\bibfnamefont {T.~C.}\ \bibnamefont
			{Killian}}, \bibinfo {author} {\bibfnamefont {S.}~\bibnamefont {Kulin}},
		\bibinfo {author} {\bibfnamefont {S.~D.}\ \bibnamefont {Bergeson}}, \bibinfo
		{author} {\bibfnamefont {L.~A.}\ \bibnamefont {Orozco}}, \bibinfo {author}
		{\bibfnamefont {C.}~\bibnamefont {Orzel}},\ and\ \bibinfo {author}
		{\bibfnamefont {S.~L.}\ \bibnamefont {Rolston}},\ }\href@noop {} {\bibfield
		{journal} {\bibinfo  {journal} {Phys. Rev. Lett.}\ }\textbf {\bibinfo
			{volume} {83}},\ \bibinfo {pages} {4776} (\bibinfo {year}
		{1999})}\BibitemShut {NoStop}%
	\bibitem [{\citenamefont {Killian}\ \emph {et~al.}(2007)\citenamefont
		{Killian}, \citenamefont {Pattard}, \citenamefont {Pohl},\ and\ \citenamefont
		{Rost}}]{kpp07}%
	\BibitemOpen
	\bibfield  {author} {\bibinfo {author} {\bibfnamefont {T.~C.}\ \bibnamefont
			{Killian}}, \bibinfo {author} {\bibfnamefont {T.}~\bibnamefont {Pattard}},
		\bibinfo {author} {\bibfnamefont {T.}~\bibnamefont {Pohl}},\ and\ \bibinfo
		{author} {\bibfnamefont {J.~M.}\ \bibnamefont {Rost}},\ }\href@noop {}
	{\bibfield  {journal} {\bibinfo  {journal} {Phys. Rep.}\ }\textbf {\bibinfo
			{volume} {449}},\ \bibinfo {pages} {77} (\bibinfo {year} {2007})}\BibitemShut
	{NoStop}%
	\bibitem [{\citenamefont {Lyon}\ and\ \citenamefont {Rolston}(2017)}]{lro17}%
	\BibitemOpen
	\bibfield  {author} {\bibinfo {author} {\bibfnamefont {M.}~\bibnamefont
			{Lyon}}\ and\ \bibinfo {author} {\bibfnamefont {S.~L.}\ \bibnamefont
			{Rolston}},\ }\href@noop {} {\bibfield  {journal} {\bibinfo  {journal} {Rep.
				Prog. Phys.}\ }\textbf {\bibinfo {volume} {80}},\ \bibinfo {pages} {017001}
		(\bibinfo {year} {2017})}\BibitemShut {NoStop}%
	\bibitem [{\citenamefont {Ichimaru}(1982)}]{ich82}%
	\BibitemOpen
	\bibfield  {author} {\bibinfo {author} {\bibfnamefont {S.}~\bibnamefont
			{Ichimaru}},\ }\href@noop {} {\bibfield  {journal} {\bibinfo  {journal} {Rev.
				Mod. Phys.}\ }\textbf {\bibinfo {volume} {54}},\ \bibinfo {pages} {1017}
		(\bibinfo {year} {1982})}\BibitemShut {NoStop}%
	\bibitem [{\citenamefont {Simien}\ \emph {et~al.}(2004)\citenamefont {Simien},
		\citenamefont {Chen}, \citenamefont {Gupta}, \citenamefont {Laha},
		\citenamefont {Martinez}, \citenamefont {Mickelson}, \citenamefont {Nagel},\
		and\ \citenamefont {Killian}}]{scg04}%
	\BibitemOpen
	\bibfield  {author} {\bibinfo {author} {\bibfnamefont {C.~E.}\ \bibnamefont
			{Simien}}, \bibinfo {author} {\bibfnamefont {Y.~C.}\ \bibnamefont {Chen}},
		\bibinfo {author} {\bibfnamefont {P.}~\bibnamefont {Gupta}}, \bibinfo
		{author} {\bibfnamefont {S.}~\bibnamefont {Laha}}, \bibinfo {author}
		{\bibfnamefont {Y.~N.}\ \bibnamefont {Martinez}}, \bibinfo {author}
		{\bibfnamefont {P.~G.}\ \bibnamefont {Mickelson}}, \bibinfo {author}
		{\bibfnamefont {S.~B.}\ \bibnamefont {Nagel}},\ and\ \bibinfo {author}
		{\bibfnamefont {T.~C.}\ \bibnamefont {Killian}},\ }\href@noop {} {\bibfield
		{journal} {\bibinfo  {journal} {Phys. Rev. Lett.}\ }\textbf {\bibinfo
			{volume} {92}},\ \bibinfo {pages} {143001} (\bibinfo {year}
		{2004})}\BibitemShut {NoStop}%
	\bibitem [{\citenamefont {Lyon}\ \emph {et~al.}(2013)\citenamefont {Lyon},
		\citenamefont {Bergeson},\ and\ \citenamefont {Murillo}}]{lbm13}%
	\BibitemOpen
	\bibfield  {author} {\bibinfo {author} {\bibfnamefont {M.}~\bibnamefont
			{Lyon}}, \bibinfo {author} {\bibfnamefont {S.~D.}\ \bibnamefont {Bergeson}},\
		and\ \bibinfo {author} {\bibfnamefont {M.~S.}\ \bibnamefont {Murillo}},\
	}\href {https://doi.org/10.1103/PhysRevE.87.033101} {\bibfield  {journal}
		{\bibinfo  {journal} {Phys. Rev. E.}\ }\textbf {\bibinfo {volume} {87}},\
		\bibinfo {pages} {033101} (\bibinfo {year} {2013})}\BibitemShut {NoStop}%
	\bibitem [{\citenamefont {Langin}\ \emph {et~al.}(2019)\citenamefont {Langin},
		\citenamefont {Gorman},\ and\ \citenamefont {Killian}}]{lgk19}%
	\BibitemOpen
	\bibfield  {author} {\bibinfo {author} {\bibfnamefont {T.~K.}\ \bibnamefont
			{Langin}}, \bibinfo {author} {\bibfnamefont {G.~M.}\ \bibnamefont {Gorman}},\
		and\ \bibinfo {author} {\bibfnamefont {T.~C.}\ \bibnamefont {Killian}},\
	}\href {https://doi.org/10.1126/science.aat3158} {\bibfield  {journal}
		{\bibinfo  {journal} {Science}\ }\textbf {\bibinfo {volume} {363}},\ \bibinfo
		{pages} {61} (\bibinfo {year} {2019})}\BibitemShut {NoStop}%
	\bibitem [{\citenamefont {Kuzmin}\ and\ \citenamefont {O'Neil}(2002)}]{kon02}%
	\BibitemOpen
	\bibfield  {author} {\bibinfo {author} {\bibfnamefont {S.~G.}\ \bibnamefont
			{Kuzmin}}\ and\ \bibinfo {author} {\bibfnamefont {T.~M.}\ \bibnamefont
			{O'Neil}},\ }\href@noop {} {\bibfield  {journal} {\bibinfo  {journal} {Phys.
				Plasmas}\ }\textbf {\bibinfo {volume} {9}},\ \bibinfo {pages} {3743}
		(\bibinfo {year} {2002})}\BibitemShut {NoStop}%
	\bibitem [{\citenamefont {Mazevet}\ \emph {et~al.}(2002)\citenamefont
		{Mazevet}, \citenamefont {Collins},\ and\ \citenamefont {Kress}}]{mck02}%
	\BibitemOpen
	\bibfield  {author} {\bibinfo {author} {\bibfnamefont {S.}~\bibnamefont
			{Mazevet}}, \bibinfo {author} {\bibfnamefont {L.~A.}\ \bibnamefont
			{Collins}},\ and\ \bibinfo {author} {\bibfnamefont {J.~D.}\ \bibnamefont
			{Kress}},\ }\href@noop {} {\bibfield  {journal} {\bibinfo  {journal} {Phys.
				Rev. Lett.}\ }\textbf {\bibinfo {volume} {88}},\ \bibinfo {pages} {55001}
		(\bibinfo {year} {2002})}\BibitemShut {NoStop}%
	\bibitem [{\citenamefont {Robicheaux}\ and\ \citenamefont
		{Hanson}(2002)}]{rha02}%
	\BibitemOpen
	\bibfield  {author} {\bibinfo {author} {\bibfnamefont {F.}~\bibnamefont
			{Robicheaux}}\ and\ \bibinfo {author} {\bibfnamefont {J.~D.}\ \bibnamefont
			{Hanson}},\ }\href@noop {} {\bibfield  {journal} {\bibinfo  {journal} {Phys.
				Rev. Lett.}\ }\textbf {\bibinfo {volume} {88}},\ \bibinfo {pages} {55002}
		(\bibinfo {year} {2002})}\BibitemShut {NoStop}%
	\bibitem [{\citenamefont {Gupta}\ \emph {et~al.}(2007)\citenamefont {Gupta},
		\citenamefont {Laha}, \citenamefont {Simien}, \citenamefont {Gao},
		\citenamefont {Castro}, \citenamefont {Killian},\ and\ \citenamefont
		{Pohl}}]{gls07}%
	\BibitemOpen
	\bibfield  {author} {\bibinfo {author} {\bibfnamefont {P.}~\bibnamefont
			{Gupta}}, \bibinfo {author} {\bibfnamefont {S.}~\bibnamefont {Laha}},
		\bibinfo {author} {\bibfnamefont {C.~E.}\ \bibnamefont {Simien}}, \bibinfo
		{author} {\bibfnamefont {H.}~\bibnamefont {Gao}}, \bibinfo {author}
		{\bibfnamefont {J.}~\bibnamefont {Castro}}, \bibinfo {author} {\bibfnamefont
			{T.~C.}\ \bibnamefont {Killian}},\ and\ \bibinfo {author} {\bibfnamefont
			{T.}~\bibnamefont {Pohl}},\ }\href@noop {} {\bibfield  {journal} {\bibinfo
			{journal} {Phys. Rev. Lett.}\ }\textbf {\bibinfo {volume} {99}},\ \bibinfo
		{pages} {75005} (\bibinfo {year} {2007})}\BibitemShut {NoStop}%
	\bibitem [{\citenamefont {Chen}\ \emph {et~al.}(2017)\citenamefont {Chen},
		\citenamefont {Witte},\ and\ \citenamefont {Roberts}}]{cwr17}%
	\BibitemOpen
	\bibfield  {author} {\bibinfo {author} {\bibfnamefont {W.~T.}\ \bibnamefont
			{Chen}}, \bibinfo {author} {\bibfnamefont {C.}~\bibnamefont {Witte}},\ and\
		\bibinfo {author} {\bibfnamefont {J.~L.}\ \bibnamefont {Roberts}},\ }\href
	{https://doi.org/10.1103/PhysRevE.96.013203} {\bibfield  {journal} {\bibinfo
			{journal} {Phys. Rev. E.}\ }\textbf {\bibinfo {volume} {96}},\ \bibinfo
		{pages} {013203} (\bibinfo {year} {2017})}\BibitemShut {NoStop}%
	\bibitem [{\citenamefont {Ott}\ and\ \citenamefont {Bonitz}(2011)}]{obo11}%
	\BibitemOpen
	\bibfield  {author} {\bibinfo {author} {\bibfnamefont {T.}~\bibnamefont
			{Ott}}\ and\ \bibinfo {author} {\bibfnamefont {M.}~\bibnamefont {Bonitz}},\
	}\href {https://doi.org/10.1103/PhysRevLett.107.135003} {\bibfield  {journal}
		{\bibinfo  {journal} {Phys. Rev. Lett.}\ }\textbf {\bibinfo {volume} {107}},\
		\bibinfo {pages} {3} (\bibinfo {year} {2011})}\BibitemShut {NoStop}%
	\bibitem [{\citenamefont {Baalrud}\ and\ \citenamefont
		{Daligault}(2017)}]{bda17a}%
	\BibitemOpen
	\bibfield  {author} {\bibinfo {author} {\bibfnamefont {S.~D.}\ \bibnamefont
			{Baalrud}}\ and\ \bibinfo {author} {\bibfnamefont {J.}~\bibnamefont
			{Daligault}},\ }\href {https://doi.org/10.1103/PhysRevE.96.043202} {\bibfield
		{journal} {\bibinfo  {journal} {Phys. Rev. E.}\ }\textbf {\bibinfo {volume}
			{96}},\ \bibinfo {pages} {043202} (\bibinfo {year} {2017})}\BibitemShut
	{NoStop}%
	\bibitem [{\citenamefont {Isaev}\ and\ \citenamefont
		{Gavriliuk}(2017)}]{iga18}%
	\BibitemOpen
	\bibfield  {author} {\bibinfo {author} {\bibfnamefont {I.~L.}\ \bibnamefont
			{Isaev}}\ and\ \bibinfo {author} {\bibfnamefont {A.~P.}\ \bibnamefont
			{Gavriliuk}},\ }\href {https://doi.org/10.1088/1361-6455/aa9b98} {\bibfield
		{journal} {\bibinfo  {journal} {J. Phys. B}\ }\textbf {\bibinfo {volume}
			{51}},\ \bibinfo {pages} {025701} (\bibinfo {year} {2018})} \BibitemShut {NoStop}%
	\bibitem [{\citenamefont {Tiwari}\ and\ \citenamefont {Baalrud}(2018)}]{tba18}%
	\BibitemOpen
	\bibfield  {author} {\bibinfo {author} {\bibfnamefont {S.~K.}\ \bibnamefont
			{Tiwari}}\ and\ \bibinfo {author} {\bibfnamefont {S.~D.}\ \bibnamefont
			{Baalrud}},\ }\href {https://doi.org/10.1063/1.5013320} {\bibfield  {journal}
		{\bibinfo  {journal} {Phys. Plasmas}\ }\textbf {\bibinfo {volume} {25}},\
		\bibinfo {pages} {013511} (\bibinfo {year} {2018})}\BibitemShut {NoStop}%
	\bibitem [{\citenamefont {Thomas}\ \emph {et~al.}(2012)\citenamefont {Thomas},
		\citenamefont {Merlino},\ and\ \citenamefont {Rosenberg}}]{tmr12}%
	\BibitemOpen
	\bibfield  {author} {\bibinfo {author} {\bibfnamefont {E.}~\bibnamefont
			{Thomas}}, \bibinfo {author} {\bibfnamefont {R.~L.}\ \bibnamefont
			{Merlino}},\ and\ \bibinfo {author} {\bibfnamefont {M.}~\bibnamefont
			{Rosenberg}},\ }\href {https://doi.org/10.1088/0741-3335/54/12/124034}
	{\bibfield  {journal} {\bibinfo  {journal} {Plasma Phys. Control Fusion}\
		}\textbf {\bibinfo {volume} {54}},\ \bibinfo {pages} {124034} (\bibinfo
		{year} {2012})}\BibitemShut {NoStop}%
	\bibitem [{\citenamefont {Karasev}\ \emph {et~al.}(2019)\citenamefont
		{Karasev}, \citenamefont {Dzlieva}, \citenamefont {D'yachkov}, \citenamefont
		{Novikov}, \citenamefont {Pavlov},\ and\ \citenamefont {Tarasov}}]{kdd19}%
	\BibitemOpen
	\bibfield  {author} {\bibinfo {author} {\bibfnamefont {V.~Y.}\ \bibnamefont
			{Karasev}}, \bibinfo {author} {\bibfnamefont {E.~S.}\ \bibnamefont
			{Dzlieva}}, \bibinfo {author} {\bibfnamefont {L.~G.}\ \bibnamefont
			{D'yachkov}}, \bibinfo {author} {\bibfnamefont {L.~A.}\ \bibnamefont
			{Novikov}}, \bibinfo {author} {\bibfnamefont {S.~I.}\ \bibnamefont
			{Pavlov}},\ and\ \bibinfo {author} {\bibfnamefont {S.~A.}\ \bibnamefont
			{Tarasov}},\ }\href {https://doi.org/10.1002/ctpp.201800136} {\bibfield
		{journal} {\bibinfo  {journal} {Cont. Plasma Phys.}\ }\textbf {\bibinfo
			{volume} {59}},\ \bibinfo {pages} {e201800136} (\bibinfo {year}
		{2019})}\BibitemShut {NoStop}%
	\bibitem [{\citenamefont {Feng}\ \emph {et~al.}(2019)\citenamefont {Feng},
		\citenamefont {Lu}, \citenamefont {Wang}, \citenamefont {Lin},\ and\
		\citenamefont {Huang}}]{flw19}%
	\BibitemOpen
	\bibfield  {author} {\bibinfo {author} {\bibfnamefont {Y.}~\bibnamefont
			{Feng}}, \bibinfo {author} {\bibfnamefont {S.}~\bibnamefont {Lu}}, \bibinfo
		{author} {\bibfnamefont {K.}~\bibnamefont {Wang}}, \bibinfo {author}
		{\bibfnamefont {W.}~\bibnamefont {Lin}},\ and\ \bibinfo {author}
		{\bibfnamefont {D.}~\bibnamefont {Huang}},\ }\href@noop {} {\bibfield
		{journal} {\bibinfo  {journal} {Rev. Mod. Phys.}\ }\textbf {\bibinfo {volume}
			{3}},\ \bibinfo {pages} {10} (\bibinfo {year} {2019})}\BibitemShut {NoStop}%
	\bibitem [{\citenamefont {Shi}\ \emph {et~al.}(2018)\citenamefont {Shi},
		\citenamefont {Qin},\ and\ \citenamefont {Fisch}}]{sqf18}%
	\BibitemOpen
	\bibfield  {author} {\bibinfo {author} {\bibfnamefont {Y.}~\bibnamefont
			{Shi}}, \bibinfo {author} {\bibfnamefont {H.}~\bibnamefont {Qin}},\ and\
		\bibinfo {author} {\bibfnamefont {N.}~\bibnamefont {Fisch}},\ }\href
	{https://doi.org/10.1063/1.5017980} {\bibfield  {journal} {\bibinfo
			{journal} {Phys. Plasmas}\ }\textbf {\bibinfo {volume} {25}},\ \bibinfo
		{pages} {055706} (\bibinfo {year} {2018})} \BibitemShut {NoStop}%
	\bibitem [{\citenamefont {Santos}\ \emph {et~al.}(2018)\citenamefont {Santos},
		\citenamefont {Bailly-Grandvaux}, \citenamefont {Ehret}, \citenamefont
		{Arefiev}, \citenamefont {Batani}, \citenamefont {Beg}, \citenamefont
		{Calisti}, \citenamefont {Ferri}, \citenamefont {Florido}, \citenamefont
		{Forestier-Colleoni}, \citenamefont {Fujioka}, \citenamefont {Gigosos},
		\citenamefont {Giuffrida}, \citenamefont {Gremillet}, \citenamefont
		{Honrubia}, \citenamefont {Kojima}, \citenamefont {Korneev}, \citenamefont
		{Law}, \citenamefont {Marqu{\`{e}}s}, \citenamefont {Morace}, \citenamefont
		{Moss{\'{e}}}, \citenamefont {Peyrusse}, \citenamefont {Rose}, \citenamefont
		{Roth}, \citenamefont {Sakata}, \citenamefont {Schaumann}, \citenamefont
		{Suzuki-Vidal}, \citenamefont {Tikhonchuk}, \citenamefont {Toncian},
		\citenamefont {Woolsey},\ and\ \citenamefont {Zhang}}]{sbe18}%
	\BibitemOpen
	\bibfield  {author} {\bibinfo {author} {\bibfnamefont {J.~J.}\ \bibnamefont
			{Santos \textit{et al.}}}}
			\href {https://doi.org/10.1063/1.5018735}
	{\bibfield  {journal} {\bibinfo  {journal} {Phys. Plasmas}\ }\textbf
		{\bibinfo {volume} {25}},\ \bibinfo {pages} {056705} (\bibinfo {year}
		{2018})}
	\BibitemShut {NoStop}%
	\bibitem [{\citenamefont {Zhang}\ \emph
		{et~al.}(2008{\natexlab{a}})\citenamefont {Zhang}, \citenamefont {Fletcher},
		\citenamefont {Rolston}, \citenamefont {Guzdar},\ and\ \citenamefont
		{Swisdak}}]{zfr08expansion}%
	\BibitemOpen
	\bibfield  {author} {\bibinfo {author} {\bibfnamefont {X.~L.}\ \bibnamefont
			{Zhang}}, \bibinfo {author} {\bibfnamefont {R.~S.}\ \bibnamefont {Fletcher}},
		\bibinfo {author} {\bibfnamefont {S.~L.}\ \bibnamefont {Rolston}}, \bibinfo
		{author} {\bibfnamefont {P.~N.}\ \bibnamefont {Guzdar}},\ and\ \bibinfo
		{author} {\bibfnamefont {M.}~\bibnamefont {Swisdak}},\ }\href@noop {}
	{\bibfield  {journal} {\bibinfo  {journal} {Phys. Rev. Lett.}\ }\textbf
		{\bibinfo {volume} {100}},\ \bibinfo {pages} {235002} (\bibinfo {year}
		{2008}{\natexlab{a}})}\BibitemShut {NoStop}%
	\bibitem [{\citenamefont {Zhang}\ \emph
		{et~al.}(2008{\natexlab{b}})\citenamefont {Zhang}, \citenamefont {Fletcher},\
		and\ \citenamefont {Rolston}}]{zfr08}%
	\BibitemOpen
	\bibfield  {author} {\bibinfo {author} {\bibfnamefont {X.~L.}\ \bibnamefont
			{Zhang}}, \bibinfo {author} {\bibfnamefont {R.~S.}\ \bibnamefont
			{Fletcher}},\ and\ \bibinfo {author} {\bibfnamefont {S.~L.}\ \bibnamefont
			{Rolston}},\ }\href@noop {} {\bibfield  {journal} {\bibinfo  {journal} {Phys.
				Rev. Lett.}\ }\textbf {\bibinfo {volume} {101}},\ \bibinfo {pages} {195002}
		(\bibinfo {year} {2008}{\natexlab{b}})}\BibitemShut {NoStop}%
	\bibitem [{\citenamefont {Dubin}\ and\ \citenamefont {O'Neil}(1999)}]{don99}%
	\BibitemOpen
	\bibfield  {author} {\bibinfo {author} {\bibfnamefont {D.~H.~E.}\
			\bibnamefont {Dubin}}\ and\ \bibinfo {author} {\bibfnamefont {T.~M.}\
			\bibnamefont {O'Neil}},\ }\href@noop {} {\bibfield  {journal} {\bibinfo
			{journal} {Rev. Mod. Phys.}\ }\textbf {\bibinfo {volume} {71}},\ \bibinfo
		{pages} {87} (\bibinfo {year} {1999})}\BibitemShut {NoStop}%
	\bibitem [{\citenamefont {Amoretti}\ \emph {et~al.}(2002)\citenamefont
		{Amoretti}, \citenamefont {Amsler}, \citenamefont {Bonomi}, \citenamefont
		{Bouchta}, \citenamefont {Bowe}, \citenamefont {Carraro}, \citenamefont
		{Cesar}, \citenamefont {Charlton}, \citenamefont {Collier}, \citenamefont
		{Doser}, \citenamefont {Filippini}, \citenamefont {Fine}, \citenamefont
		{Fontana}, \citenamefont {Fujiwara}, \citenamefont {Funakoshi}, \citenamefont
		{Genova}, \citenamefont {Hangst}, \citenamefont {Hayano}, \citenamefont
		{Holzscheiter}, \citenamefont {Jorgensen}, \citenamefont {Lagomarsino},
		\citenamefont {Landua}, \citenamefont {Lindelof}, \citenamefont {Rizzini},
		\citenamefont {Madsen}, \citenamefont {Manuzio}, \citenamefont {Marchesotti},
		\citenamefont {Montagna}, \citenamefont {Pruys}, \citenamefont {Regenfus},
		\citenamefont {Riedler}, \citenamefont {Rochet}, \citenamefont {Rotondi},
		\citenamefont {Rouleau}, \citenamefont {Testera}, \citenamefont {Variola},
		\citenamefont {Watson},\ and\ \citenamefont {van~der Werf}}]{aab02}%
	\BibitemOpen
	\bibfield  {author} {\bibinfo {author} {\bibfnamefont {M.}~\bibnamefont
			{Amoretti \textit{et al.}}}}
			\href@noop {} {\bibfield  {journal} {\bibinfo
			{journal} {{Nature}}\ }\textbf {\bibinfo {volume} {419}},\ \bibinfo {pages}
		{456} (\bibinfo {year} {2002})}\BibitemShut {NoStop}%
	\bibitem [{\citenamefont {Gabrielse}\ \emph {et~al.}(2002)\citenamefont
		{Gabrielse}, \citenamefont {Bowden}, \citenamefont {Oxley}, \citenamefont
		{Speck}, \citenamefont {Storry}, \citenamefont {Tan}, \citenamefont
		{Wessels}, \citenamefont {Grzonka}, \citenamefont {Oelert}, \citenamefont
		{Schepers}, \citenamefont {Sefzick}, \citenamefont {Walz}, \citenamefont
		{Pittner}, \citenamefont {Hansch},\ and\ \citenamefont
		{Collaboration}}]{gbo02}%
	\BibitemOpen
	\bibfield  {author} {\bibinfo {author} {\bibfnamefont {G.}~\bibnamefont
			{Gabrielse \textit{et al.}}}}
			\href@noop {} {\bibfield  {journal} {\bibinfo  {journal}
			{Phys. Rev. Lett.}\ }\textbf {\bibinfo {volume} {89}},\ \bibinfo {pages}
		{213401} (\bibinfo {year} {2002})}\BibitemShut {NoStop}%
	\bibitem [{\citenamefont {Choi}\ \emph {et~al.}(2008)\citenamefont {Choi},
		\citenamefont {Knuffman}, \citenamefont {Zhang}, \citenamefont {Povilus},\
		and\ \citenamefont {Raithel}}]{ckz08}%
	\BibitemOpen
	\bibfield  {author} {\bibinfo {author} {\bibfnamefont {J.~H.}\ \bibnamefont
			{Choi}}, \bibinfo {author} {\bibfnamefont {B.}~\bibnamefont {Knuffman}},
		\bibinfo {author} {\bibfnamefont {X.~H.}\ \bibnamefont {Zhang}}, \bibinfo
		{author} {\bibfnamefont {A.~P.}\ \bibnamefont {Povilus}},\ and\ \bibinfo
		{author} {\bibfnamefont {G.}~\bibnamefont {Raithel}},\ }\href
	{https://doi.org/10.1103/PhysRevLett.100.175002} {\bibfield  {journal}
		{\bibinfo  {journal} {Phys. Rev. Lett.}\ }\textbf {\bibinfo {volume} {100}},\
		\bibinfo {pages} {175002} (\bibinfo {year} {2008})}\BibitemShut {NoStop}%
	\bibitem [{\citenamefont {Nagel}\ \emph {et~al.}(2003)\citenamefont {Nagel},
		\citenamefont {Simien}, \citenamefont {Laha}, \citenamefont {Gupta},
		\citenamefont {Ashoka},\ and\ \citenamefont {Killian}}]{nsl03}%
	\BibitemOpen
	\bibfield  {author} {\bibinfo {author} {\bibfnamefont {S.~B.}\ \bibnamefont
			{Nagel}}, \bibinfo {author} {\bibfnamefont {C.~E.}\ \bibnamefont {Simien}},
		\bibinfo {author} {\bibfnamefont {S.}~\bibnamefont {Laha}}, \bibinfo {author}
		{\bibfnamefont {P.}~\bibnamefont {Gupta}}, \bibinfo {author} {\bibfnamefont
			{V.~S.}\ \bibnamefont {Ashoka}},\ and\ \bibinfo {author} {\bibfnamefont
			{T.~C.}\ \bibnamefont {Killian}},\ }\href@noop {} {\bibfield  {journal}
		{\bibinfo  {journal} {Phys. Rev. A.}\ }\textbf {\bibinfo {volume} {67}},\
		\bibinfo {eid} {011401(R)} (\bibinfo {year} {2003})}\BibitemShut {NoStop}%
	\bibitem [{\citenamefont {Murillo}(2001)}]{mur01}%
	\BibitemOpen
	\bibfield  {author} {\bibinfo {author} {\bibfnamefont {M.~S.}\ \bibnamefont
			{Murillo}},\ }\href@noop {} {\bibfield  {journal} {\bibinfo  {journal} {Phys.
				Rev. Lett.}\ }\textbf {\bibinfo {volume} {87}},\ \bibinfo {pages} {115003}
		(\bibinfo {year} {2001})}\BibitemShut {NoStop}%
	\bibitem [{\citenamefont {Castro}\ \emph {et~al.}(2008)\citenamefont {Castro},
		\citenamefont {Gao},\ and\ \citenamefont {Killian}}]{cgk08}%
	\BibitemOpen
	\bibfield  {author} {\bibinfo {author} {\bibfnamefont {J.}~\bibnamefont
			{Castro}}, \bibinfo {author} {\bibfnamefont {H.}~\bibnamefont {Gao}},\ and\
		\bibinfo {author} {\bibfnamefont {T.~C.}\ \bibnamefont {Killian}},\
	}\href@noop {} {\bibfield  {journal} {\bibinfo  {journal} {Plasma Phys.
				Control Fusion}\ }\textbf {\bibinfo {volume} {50}},\ \bibinfo {pages}
		{124011} (\bibinfo {year} {2008})}\BibitemShut {NoStop}%
	\bibitem [{\citenamefont {Kulin}\ \emph {et~al.}(2000)\citenamefont {Kulin},
		\citenamefont {Killian}, \citenamefont {Bergeson},\ and\ \citenamefont
		{Rolston}}]{kkb00}%
	\BibitemOpen
	\bibfield  {author} {\bibinfo {author} {\bibfnamefont {S.}~\bibnamefont
			{Kulin}}, \bibinfo {author} {\bibfnamefont {T.~C.}\ \bibnamefont {Killian}},
		\bibinfo {author} {\bibfnamefont {S.~D.}\ \bibnamefont {Bergeson}},\ and\
		\bibinfo {author} {\bibfnamefont {S.~L.}\ \bibnamefont {Rolston}},\
	}\href@noop {} {\bibfield  {journal} {\bibinfo  {journal} {Phys. Rev. Lett.}\
		}\textbf {\bibinfo {volume} {85}},\ \bibinfo {pages} {318} (\bibinfo {year}
		{2000})}\BibitemShut {NoStop}%
	\bibitem [{\citenamefont {Laha}\ \emph {et~al.}(2007)\citenamefont {Laha},
		\citenamefont {Gupta}, \citenamefont {Simien}, \citenamefont {Gao},
		\citenamefont {Castro},\ and\ \citenamefont {Killian}}]{lgs07}%
	\BibitemOpen
	\bibfield  {author} {\bibinfo {author} {\bibfnamefont {S.}~\bibnamefont
			{Laha}}, \bibinfo {author} {\bibfnamefont {P.}~\bibnamefont {Gupta}},
		\bibinfo {author} {\bibfnamefont {C.~E.}\ \bibnamefont {Simien}}, \bibinfo
		{author} {\bibfnamefont {H.}~\bibnamefont {Gao}}, \bibinfo {author}
		{\bibfnamefont {J.}~\bibnamefont {Castro}},\ and\ \bibinfo {author}
		{\bibfnamefont {T.~C.}\ \bibnamefont {Killian}},\ }\href@noop {} {\bibfield
		{journal} {\bibinfo  {journal} {Phys. Rev. Lett.}\ }\textbf {\bibinfo
			{volume} {99}},\ \bibinfo {pages} {155001} (\bibinfo {year}
		{2007})}\BibitemShut {NoStop}%
	\bibitem [{\citenamefont {Warrens}\ \emph {et~al.}(2020)\citenamefont
		{Warrens}, \citenamefont {Gorman},\ and\ \citenamefont {Killian}}]{wgk20}%
	\BibitemOpen
	\bibfield  {author} {\bibinfo {author} {\bibfnamefont {M.~K.}\ \bibnamefont
			{Warrens}}, \bibinfo {author} {\bibfnamefont {G.~M.}\ \bibnamefont
			{Gorman}},\ and\ \bibinfo {author} {\bibfnamefont {T.~C.}\ \bibnamefont
			{Killian}}} (\bibinfo {year} {2020}),\ \bibinfo {note} {unpublished
		Manuscript}\BibitemShut {NoStop}%
	\bibitem [{\citenamefont {Baitin}\ and\ \citenamefont
		{Kuzanyan}(1998)}]{bku98}%
	\BibitemOpen
	\bibfield  {author} {\bibinfo {author} {\bibfnamefont {A.~V.}\ \bibnamefont
			{Baitin}}\ and\ \bibinfo {author} {\bibfnamefont {K.~M.}\ \bibnamefont
			{Kuzanyan}},\ }\href@noop {} {\bibfield  {journal} {\bibinfo  {journal} {J.
				Plasma Phys.}\ }\textbf {\bibinfo {volume} {59}},\ \bibinfo {pages} {83}
		(\bibinfo {year} {1998})}\BibitemShut {NoStop}%
	\bibitem [{\citenamefont {Dorozhkina}\ and\ \citenamefont
		{Semenov}(1998)}]{dse98}%
	\BibitemOpen
	\bibfield  {author} {\bibinfo {author} {\bibfnamefont {D.~S.}\ \bibnamefont
			{Dorozhkina}}\ and\ \bibinfo {author} {\bibfnamefont {V.~E.}\ \bibnamefont
			{Semenov}},\ }\href@noop {} {\bibfield  {journal} {\bibinfo  {journal} {Phys.
				Rev. Lett.}\ }\textbf {\bibinfo {volume} {81}},\ \bibinfo {pages} {2691}
		(\bibinfo {year} {1998})}\BibitemShut {NoStop}%
	\bibitem [{\citenamefont {Bosch}\ and\ \citenamefont {Merlino}(1986)}]{bme86c}%
	\BibitemOpen
	\bibfield  {author} {\bibinfo {author} {\bibfnamefont {R.~A.}\ \bibnamefont
			{Bosch}}\ and\ \bibinfo {author} {\bibfnamefont {R.~L.}\ \bibnamefont
			{Merlino}},\ }\href {https://doi.org/10.1063/1.865628} {\bibfield  {journal}
		{\bibinfo  {journal} {Phys. Fluids}\ }\textbf {\bibinfo {volume} {29}},\
		\bibinfo {pages} {1998} (\bibinfo {year} {1986})}\BibitemShut {NoStop}%
	\bibitem [{\citenamefont {Bergeson}\ \emph {et~al.}(2019)\citenamefont
		{Bergeson}, \citenamefont {Baalrud}, \citenamefont {{Leland Ellison}},
		\citenamefont {Grant}, \citenamefont {Graziani}, \citenamefont {Killian},
		\citenamefont {Murillo}, \citenamefont {Roberts},\ and\ \citenamefont
		{Stanton}}]{bbl19}%
	\BibitemOpen
	\bibfield  {author} {\bibinfo {author} {\bibfnamefont {S.~D.}\ \bibnamefont
			{Bergeson}}, \bibinfo {author} {\bibfnamefont {S.~D.}\ \bibnamefont
			{Baalrud}}, \bibinfo {author} {\bibfnamefont {C.}~\bibnamefont {{Leland
					Ellison}}}, \bibinfo {author} {\bibfnamefont {E.}~\bibnamefont {Grant}},
		\bibinfo {author} {\bibfnamefont {F.~R.}\ \bibnamefont {Graziani}}, \bibinfo
		{author} {\bibfnamefont {T.~C.}\ \bibnamefont {Killian}}, \bibinfo {author}
		{\bibfnamefont {M.~S.}\ \bibnamefont {Murillo}}, \bibinfo {author}
		{\bibfnamefont {J.~L.}\ \bibnamefont {Roberts}},\ and\ \bibinfo {author}
		{\bibfnamefont {L.~G.}\ \bibnamefont {Stanton}},\ }\href
	{https://doi.org/10.1063/1.5119144} {\bibfield  {journal} {\bibinfo
			{journal} {Phys. Plasmas}\ }\textbf {\bibinfo {volume} {26}},\ \bibinfo
		{pages} {100501} (\bibinfo {year} {2019})}\BibitemShut {NoStop}%
\end{thebibliography}
%
\end{document}